\documentclass[12pt]{article}
\usepackage{epsfig}
\usepackage{subfigure}
\usepackage{amsmath}
\usepackage{amsfonts}
\usepackage{amssymb}
\usepackage{url}
\usepackage{hyperref}
\newcommand{\be}{\begin{equation}}
\newcommand{\ee}{\end{equation}}
\newcommand{\bea}{\begin{eqnarray}}
\newcommand{\eea}{\end{eqnarray}}

\newcommand{\Slash}[1]{\ooalign{\hfil/\hfil\crcr$#1$}}
\begin{document}
\begin{titlepage}
\begin{flushright}
February 27, 2008
\end{flushright}
\vspace{4\baselineskip}
\begin{center}
{\large\bf 
Constraint on the heavy sterile neutrino mixing angles\\
in the SO(10) model with double see-saw mechanism}
\end{center}
\vspace{1cm}
\begin{center}
{\large
Takeshi Fukuyama$^{a}$,
\footnote{\tt E-mail:fukuyama@se.ritsumei.ac.jp}
Tatsuru Kikuchi$^{b,c}$
\footnote{\tt E-mail:tatsuru@post.kek.jp}
and Koichi Matsuda$^{d}$
\footnote{\tt E-mail:matsuda@mail.tsinghua.edu.cn}
}
\end{center}
\vspace{0.2cm}
\begin{center}
${}^{a}$ 
{\small \it Department of Physics, Ritsumeikan University,
Kusatsu, Shiga 525-8577, Japan}\\
${}^{b}$ 
{\small \it Theory Division, KEK,
Oho 1-1, Tsukuba, Ibaraki 305-0801, Japan}\\
${}^{c}$ {\small \it
Department of Physics, Oklahoma State University, Stillwater, OK 74078, USA}\\
${}^{d}$ 
{\small \it Center for High Energy Physics, Tsinghua University
Beijing 100084, China}\\
\medskip
\vskip 5mm
\end{center}
\vskip 5mm
%
%
\begin{abstract}
Constraints on the heavy sterile neutrino mixing angles are studied
in the framework of a minimal supersymmetric SO(10) model
with the use of the double see-saw mechanism. 
A new singlet matter in addition to the right-handed neutrinos is introduced 
to realize the double see-saw mechanism. 
The light Majorana neutrino mass matrix is, in general, given by 
a combination of those of the singlet neutrinos and the active neutrinos.
The minimal SO(10) model is used to give an example form of
the Dirac neutrino mass matrix, which enables us to predict the masses
and  the mixing angles in the enlarged $9 \times 9$ neutrino mass matrix. 
Mixing angles between the light Majorana neutrinos and the heavy sterile 
neutrinos are shown to be within the LEP experimental bound on 
all ranges of the Majorana phases.
\end{abstract}
\end{titlepage}
\section{Introduction}
Recent neutrino oscillation data opened up a new window to prove physics 
beyond the Standard Model. 
As pointed out in \cite{Weinberg:1979sa}, we can construct, 
within the context of the standard model (SM), an operator 
which gives rise to the neutrino masses as 
\be
{\cal L}_{\rm eff} = \frac{1}{\Lambda} \,(\ell_L H)^{T} C^{-1} (\ell_L H)\;.
\ee
Here $\ell$, $H$ are the lepton doublet and the Higgs doublet, 
$C$ is the charge conjugation operator and $\Lambda$ is the scale 
in which something new physics appears. This operator can naturally 
be arisen in the see-saw mechanism \cite{see-saw}, which may give 
a guideline to construct models of new physics through the existence of 
the right-handed neutrinos. 

On the other hand, the supersymmetric (SUSY) grand unified theory (GUT) provides 
an attractive implication for the understandings of the low-energy 
physics. In fact, for instance, the anomaly cancellation between 
the several matter multiplets is automatic in the GUT based on 
a simple gauge group, since the matter multiplets are unified into 
a few multiplets, the experimental data supports the fact of unification 
of three gauge couplings at the GUT scale 
$M_{\rm GUT} = 2 \times 10^{16}$ [GeV] 
assuming the particle contents of the minimal supersymmetric 
standard model (MSSM), and also the right-handed 
neutrino appeared naturally in the SO(10) GUT provides a natural 
explanation of the smallness of the neutrino masses through 
the see-saw mechanism \cite{see-saw}.

Although the essential concept of the see-saw mechanism is the same, 
there can be many possibilities according to the types of the see-saw mechanism. 
For instance, as motivated by the superstring inspired $E_6$ models, 
we come to consider {\it the double see-saw mechanism} 
\cite{EW86, Mohapatra:1986aw, Mohapatra:1986bd, Fukuyama:2005gg, 
Malinsky:2005bi} and it's extension, the type-III see-saw mechanism \cite{Akhmedov:1995vm}
(see also, \cite{Barr:2003nn, Barr:2005ss}).
Interestingly, in such an extension of the standard see-saw mechanism, 
it may appear the singlet neutrinos in the reachable range of the future collider experiments.
The possibility of testing the not-so-heavy singlet neutrinos at collider experiments
has firstly been proposed by \cite{Dittmar:1989yg} and subsequently analyzed by
the LEP collaborations \cite{Achard:2001qv}.

In this letter we give constraints on the mixing angles between 
active and sterile neutrinos in the enlarged $9 \times 9$ mass matrix which appears
in the double see-saw mechanism using an SO(10) model with double see-saw mechanism.
The constraints on the mixing angles are imposed so as to satisfy the current
neutrino oscillation data.

We accept the same Lagrangian as in \cite{Fukuyama:2005gg}. 
That is, we add a new singlet matter ($S$) in addition to the right-handed 
neutrino ($\nu^c$) per a generation. 
The Lagrangian in this model is given by
\be
{\cal L}_Y = Y_\nu^{ij} ~\nu^c_i L_j \, H_u 
+ Y_s^{ij} ~\nu^c_i S_j \, H_s
+ \mu_s ~S_i^2 + h.c. \;,
\label{1}
\ee
where $L_j$ is the lepton doublet, and $H_u$, $H_s$ are the $SU(2)_L$ 
doublet, singlet Higgs fields. 

Note that the $\mu_s$ term in the above breaks an originally existing 
global ${\rm U}(1)_L$ (Lepton Number) (and ${\rm U}(1)_{\cal R}$ symmetry in the case of supersymmetry). 
Thus we can naturally expect it as a small value compared with the electroweak scale 
even around the keV scale, according to the following reason: 
when the $\mu_s$ term is arisen from the VEV of a singlet 
$\mu_s = \lambda \left<S' \right>$, there appears a pseudo-NG boson, 
called Majoron $J=\Im S'$ associated with the spontaneously broken 
${\rm U}(1)_L$ symmetry. Then the keV scale lepton number violation may lead 
to an interesting signature in the neutrinoless double beta decay 
\cite{Berezhiani:1992cd} or becomes a possible candidate for 
the cold dark matter \cite{Berezinsky:1993fm}.

The mass terms of the Lagrangian (\ref{1}) are re-written in a matrix form 
in the base with $\{\nu,\,\nu^c,\,S \}$ as follows \cite{EW86, 
Mohapatra:1986aw, Mohapatra:1986bd, Fukuyama:2005gg, Malinsky:2005bi},
\be
{\cal M}
=\left(
\begin{array}{ccc}
0    & m_D  &  0   \\
m_D^T  & 0      &  M_D \\
0    & M_D^T    & \mu_s \,\mathbb{I}
\end{array}
\right)\;.
\label{2}
\ee
Here $m_D \equiv Y_\nu \left<H_u \right>$, $M_D \equiv Y_s \left<H_s \right>$, 
and $\mathbb{I} \equiv {\rm diag}(1,1,1)$. In this paper we assume that 
the mass matrix $M_D$ is written in terms of a unitary matrix 
$V$ as $(M_D)_{ij} = V_{ij}^* M_{Dj}$, where the unitary matrix $V$ 
diagonalises a combination $M_D M_D^T$,
\be
V^T~ M_D M_D^T~ V =
{\rm diag} \left(M_{D1}^2, M_{D2}^2, M_{D3}^2 \right) \;.
\label{MD}
\ee
On the other hand, the full $9 \times 9$ mass matrix (\ref{2}) 
can be diagonalised by using a unitary matrix $U$ as
\be
U^{T} {\cal M} ~U = {\rm diag}(m_1,m_2,m_3, 
\underbrace{m_{N_1}, m_{N_2}, \cdots, 
m_{N_5}, m_{N_6}}_{\mbox{heavy isosinglet neutrinos}})\;,
\ee
where $m_{N_1} \simeq m_{N_2} < m_{N_3} \simeq m_{N_4} 
< m_{N_5} \simeq m_{N_6}$. 
If the eigenvalues of each $3\times 3$ matrix satisfy 
$\mu_s \ll m_{Di} \ll M_{Di}$ as was assumed in \cite{Fukuyama:2005gg},
the light mass eigenvalue is roughly given by
$M_\nu \sim \mu_s (m_{D}/M_{D})^2$.
The MNS mixing matrix $U_{\rm MNS}$ is the first $3 \times 3$ part of 
this unitary matrix $U$,
\be
U
=\left(
\begin{array}{cc}
U_{\rm MNS}  & 
\begin{array}{c}
U_{e A} \\
U_{\mu A} \\
U_{\tau A}
\end{array}
\\
* & *
\end{array}
\right)\;.
\ee
Here the label $A$ runs over the extra mass eigenstates $A=4,\cdots,9$,
and the extraordinary matrix element $U_{eA}$ gives a sterile to active 
neutrino mixing angle that have to be small enough so as to satisfy 
the current experimental bound, which is obtained from the invisible decays 
of the $Z$ boson measured in L3 experiment at LEP.

After integrating out the heavy singlets, $\nu^c$ and $S$,
we obtain the effective light neutrino mass matrix as 
\be
M_\nu =  \left(M_D^{-1} m_D \right)^{T} \mu_s 
\left(M_D^{-1} m_D \right)\;.
\label{6}
\ee
This light Majorana mass matrix can be diagonalised by the MNS matrix,
\be
U_{\rm MNS}^{T} ~M_\nu ~U_{\rm MNS} = {\rm diag} \left(m_1,m_2,m_3 \right)\;.
\ee
An important fact is that the new physics scale has also the 
``see-saw structure'' as 
\be
\Lambda \cong \frac{M_D^2}{\mu_s}\;.
\ee
Hence this mechanism is sometimes called as ``double see-saw'' mechanism. 
It's not the actual see-saw type but the inverse see-saw form, because 
the small lepton number violating ($\Slash{L}$) scale $\mu_s$ 
would indicate the large scale. 

\section{Fermion masses in an SO(10) Model with a singlet}
In order to make a prediction on the second Dirac neutrino mass matrix 
$M_D$, we need an information for the Yukawa couplings of $Y_\nu$. 
In this paper, we make the minimal SO(10) model extend to add 
a number of singlet, which preserves a precise information for $m_D$.

Now we give a brief review of the minimal SUSY SO(10) model proposed in 
\cite{Babu:1992ia} and recently analysed in detail in references 
\cite{Matsuda:2000zp, Matsuda:2001bg, Fukuyama:2002ch, Bajc:2002iw, 
Goh:2003sy, Goh:2003hf, Dutta:2004wv, Matsuda:2004bq, Bertolini:2005qb, Babu:2005ia}. 
Even when we concentrate our discussion on the issue of how to reproduce the 
realistic fermion mass matrices in the SO(10) model, there are lots of 
possibilities of the introduction of Higgs multiplets. 
The minimal supersymmetric SO(10) model includes only one {\bf 10} 
and one $\overline{\bf 126}$ Higgs multiplets in Yukawa couplings 
with {\bf 16} matter multiplets. 
Here, in addition to it, we introduce a number of SO(10) singlet 
chiral superfields ${\bf 1}$ as new matter multiplets 
\footnote{The singlet matter multiplet may have it's origin in some $E_6$ 
representations ${\bf 27}$ or ${\bf 78}$ which are decomposed under the SO(10) 
subgroup as ${\bf 27} = {\bf 16 + 10 +1}$, 
${\bf 78} = {\bf 45 + 16 + \overline{16} + 1}$. 
In such a case, the superpotential given in Eq. (\ref{WY}) may be generated 
from the following $E_6$ invariant superpotential: 
$W_Y = Y_{1}^{ij} {\bf 27}_i {\bf 27}_j {\bf 27}_H 
+ Y_{2}^{ij} {\bf 27}_i {\bf 27}_j {\bf 351}^\prime_H 
+ Y_{3}^{ij} {\bf 27}_i {\bf 78}_j \overline{{\bf 27}}_H
+ \mu^{ij} {\bf 27}_i {\bf 27}_j$. }. 
This additional singlet can provide a double see-saw mechanism as 
described in the previous section. The relevant superpotential can be 
written as 
\be
W_Y = Y_{10}^{ij} {\bf 16}_i {\bf 16}_j {\bf 10}_H
+ Y_{126}^{ij} {\bf 16}_i {\bf 16}_j \overline{{\bf 126}}_H
+ Y_{s}^{ij} {\bf 16}_i {\bf 1}_j \overline{{\bf 16}}_H
+ \mu_s {\bf 1}_i^2 \;.
\label{WY}
\ee
At low energy after the GUT symmetry breaking, the superpotential leads to 
\begin{eqnarray}
W &=&
\left(Y_{10}^{ij} H_{10}^u + Y_{126}^{ij} H_{126}^u \right) u^c_i q_j
+
\left(Y_{10}^{ij} H_{10}^d + Y_{126}^{ij} H_{126}^d \right) d^c_i q_j
\nonumber\\
&+&
\left(Y_{10}^{ij} H_{10}^u -3  Y_{126}^{ij} H_{126}^u \right) N_i {\ell}_j
+
\left(Y_{10}^{ij} H_{10}^d -3 Y_{126}^{ij} H_{126}^d \right) e^c_i {\ell}_j
\nonumber\\
&+& Y_s^{ij} N_i S_j H_s + \mu_s S_i^2 \;, 
\end{eqnarray}
where $H_{10}$ and $H_{126}$ correspond to the Higgs doublets in ${\bf 10}_H$ 
and $\overline{{\bf 126}}_H$. That is, we have two pairs of Higgs doublets. 
In order to keep the successful gauge coupling unification, 
we suppose that one pair of Higgs doublets 
(a linear combination of $H_{10}^{u,d}$ and $H_{126}^{u,d}$) 
is light while the other pair is  heavy ($\simeq M_{\rm GUT}$). 
The light Higgs doublets are identified 
as the MSSM Higgs doublets ($H_u$ and $H_d$) and given by 
\be
H_u \ =\ \widetilde{\alpha}_u ~H_{10}^u + \widetilde{\beta}_u ~H_{126}^u \;;
\quad
H_d \ =\ \widetilde{\alpha}_d ~H_{10}^d + \widetilde{\beta}_d ~H_{126}^d \;,
\label{mix}
\ee
where $\widetilde{\alpha}_{u,d}$ and $\widetilde{\beta}_{u,d}$ denote 
elements of the unitary matrix which rotate the flavour basis in the original 
model into the SUSY mass eigenstates. Omitting the heavy Higgs mass 
eigenstates, the low energy superpotential is described by only the light 
Higgs doublets $H_u$ and $H_d$ such that 
\begin{eqnarray}
W_Y &=&
\left( \alpha^u  Y_{10}^{ij} + \beta^u  Y_{126}^{ij} \right) u^c_i q_j H_u
\ +\
\left( \alpha^d  Y_{10}^{ij} + \beta^d  Y_{126}^{ij} \right) d^c_i q_j H_d
\nonumber\\
&+&
\left( \alpha^u  Y_{10}^{ij} -3 \beta^u Y_{126}^{ij} \right) N_i \ell_j H_u
\ +\
\left( \alpha^d  Y_{10}^{ij} -3 \beta^d Y_{126}^{ij} \right) e_i^c \ell_j H_d
\nonumber\\
&+& Y_s^{ij} N_i S_j H_s \ +\ \mu_s S_i^2 \;,
\label{Yukawa3}
\end{eqnarray} 
where the formulas of the inverse unitary transformation of Eq.~(\ref{mix}), 
$H_{10}^{u,d} = \alpha^{u,d} H_{u,d} + \cdots $ and 
$H_{126}^{u,d} = \beta^{u,d} H_{u,d} + \cdots $, have been used. 
Providing the Higgs VEV's, $\langle H_u \rangle = v \sin \beta$ and 
$\langle H_d \rangle = v \cos \beta$ with $v \simeq 174$ [GeV], 
the Dirac mass matrices can be read off as 
\begin{eqnarray}
M_u &=& c_{10} M_{10} + c_{126} M_{126}, 
\nonumber \\
M_d &=& M_{10} + M_{126},   
\nonumber \\
m_D &=& c_{10} M_{10} - 3 c_{126} M_{126}, 
\nonumber \\
M_e &=& M_{10} - 3 M_{126}, 
\label{massmatrix}
\end{eqnarray} 
where $M_u$, $M_d$, $m_D$ and $M_e$ denote up-type quark, down-type quark, 
Dirac neutrino and charged-lepton mass matrices, respectively. Note that 
all the quark and lepton mass matrices are characterised by only two basic 
mass matrices, $M_{10}$ and $M_{126}$, and four complex coefficients 
$c_{10}$ and $c_{126}$. In addition to the above mass matrices the above 
model indicates the mass matrices, 
\begin{eqnarray}
M_R &=& c_R~M_{126}\;,
\nonumber\\
M_L &=& c_L~M_{126}\;,
\label{massmatrix2}
\end{eqnarray} 
together with $M_D$ given in Eq. (\ref{MD}). 
$c_R$ and $c_L$ correspond to the VEV's of 
$({\bf 10}, {\bf 1}, {\bf 3}) \subset {\overline{\bf126}}$ 
and $({\overline{\bf 10}}, {\bf 3}, {\bf 1}) \subset {\overline{\bf126}}$
under the the Pati-Salam subgroup, $G_{422}=SU(4)_c \times SU(2)_L \times SU(2)_R$. 

If $M_R$, $M_L$, $M_D$ terms dominate, 
they are called Type-I, Type-II, and double see-saw, respectively. 
In this paper, we consider the case $c_R \ =\ c_L \ =\ 0 $, double.

The mass matrix formulas in Eq.~(\ref{massmatrix}) leads to the GUT 
relation among the quark and lepton mass matrices, 
\begin{eqnarray}
M_e = c_d \left( M_d + \kappa  M_u \right) \; , 
\label{GUTrelation} 
\end{eqnarray} 
where 
\begin{eqnarray}
c_d &=& - \frac{3 c_{10} + c_{126}}{c_{10}-c_{126}}, 
\\
\kappa &=& - \frac{4}{3 c_{10} + c_{126}}. 
\end{eqnarray} 
Without loss of generality, we can take the basis where $M_u$ is real 
and diagonal, $M_u = D_u$. Since $M_d$ is the symmetric matrix, it is 
described as $M_d = V_{\mathrm{CKM}}^* \,D_d \,V_{\mathrm{CKM}}^\dagger$ 
by using the CKM matrix $V_{\mathrm{CKM}}$ and the real diagonal mass matrix 
$D_d$. 
Considering the basis-independent quantities, $\mathrm{tr}[M_e^\dagger M_e ]$, 
$\mathrm{tr} [(M_e^\dagger M_e)^2 ]$ and $\mathrm{det} [M_e^\dagger M_e ]$, 
and eliminating $|c_d|$, we obtain two independent equations,  
\begin{eqnarray}
\left(
\frac{\mathrm{tr} [\widetilde{M_e}^\dagger \widetilde{M_e} ]}
{m_e^2 + m_{\mu}^2 + m_{\tau}^2} \right)^2
&=& 
\frac{\mathrm{tr} [( \widetilde{M_e}^\dagger \widetilde{M_e} )^2 ]}
{m_e^4 + m_{\mu}^4 + m_{\tau}^4},
\label{cond1} \\ 
\left( \frac{\mathrm{tr} [\widetilde{M_e}^\dagger \widetilde{M_e} ]}
{m_e^2 + m_{\mu}^2 + m_{\tau}^2} \right)^3
&=&
\frac{\mathrm{det} [\widetilde{M_e}^\dagger \widetilde{M_e} ]}
{m_e^2 \; m_\mu^2 \; m_\tau^2},
\label{cond2} 
\end{eqnarray}
where $\widetilde{M_e} \equiv 
V_{\mathrm{CKM}}^* \, D_d \, V_{\mathrm{CKM}}^\dagger + \kappa D_u$. 
With input data of six quark masses, three angles and one CP-phase in the 
CKM matrix and three charged-lepton masses, we can solve the above equations 
and determine $\kappa$ and $|c_d|$, but one parameter, the phase of $c_d$, 
is left undetermined \cite{Matsuda:2000zp, Matsuda:2001bg, Fukuyama:2002ch}. 
With input data of six quark masses, three angles and one CP-phase 
in the CKM matrix and three charged lepton masses, 
we solve the above equations and determine $\kappa$. 
The original basic mass matrices, $M_{10}$ and $M_{126}$, are described by 
\begin{eqnarray}
M_{10} 
&=& 
\frac{3+ |c_d| e^{i \sigma}}{4} 
V_{\mathrm{CKM}}^* \, D_d \, V_{\mathrm{CKM}}^\dagger
+ \frac{|c_d| e^{i \sigma} \kappa}{4} D_u, 
\label{M10}
\\
M_{126} &=& 
\frac{1- |c_d| e^{i \sigma}}{4} 
V_{\mathrm{CKM}}^* \, D_d \, V_{\mathrm{CKM}}^\dagger
-\frac{|c_d| e^{i \sigma} \kappa}{4} D_u, 
\label{M126} 
\end{eqnarray}
as the functions of $\sigma$, the phase of $c_d$, with the solutions 
$|c_d|$ and $\kappa$ determined by the GUT relation.  

Now let us solve the GUT relation and determine $|c_d|$ and $\kappa$. 
Since the GUT relation of Eq.~(\ref{GUTrelation}) is valid only at the GUT 
scale, we first evolve the data at the weak scale to the corresponding 
quantities at the GUT scale with given $\tan \beta$ according to the 
renormalization group equations (RGE's) and use them as input data at the 
GUT scale. Note that it is non-trivial to find the solution of the GUT 
relation since the number of the free parameters (fourteen) is almost the 
same as the number of inputs (thirteen). The solution of the GUT relation 
exists only if we take appropriate input parameters. Taking the experimental 
data at the $M_Z$ scale \cite{FK}, we get the following values 
for charged fermion masses and the CKM matrix at the GUT scale, $M_{\rm GUT}$ 
with $\tan \beta = 10$: 
\begin{eqnarray}
& & m_u = 0.000980 \; , \; \; m_c = 0.285 \; , \; \;  m_t = 113, 
\nonumber\\
& & m_d = 0.00135 \; , \; \; m_s = 0.0201 \; , \; \; m_b = 0.996, 
\nonumber\\ 
& & m_e = 0.000326 \; , \; \; m_\mu = 0.0687 \; , \; \; m_\tau = 1.17, 
\nonumber
\end{eqnarray}
and 
\begin{eqnarray}
V_{\mathrm{CKM}}(M_{\rm GUT}) 
= \left( 
\begin{array}{ccc}
0.975 & 0.222 & - 0.000940 - 0.00289 i \\
-0.222 - 0.000129 i & 0.974 + 0.000124 i & 0.0347 \\ 
0.00864 - 0.00282 i & - 0.0337 - 0.000647 i & 0.999
\end{array} 
\right) \;
\nonumber
\end{eqnarray}
in the standard parameterisation. The signs of the input fermion masses 
have been chosen to be $(m_u, m_c, m_t) = (+, -, +)$ and 
$(m_d, m_s, m_b) = (-, -, +)$. 
By using these outputs at the GUT scale as input parameters, 
we can solve Eqs.~(\ref{cond1}) and (\ref{cond2}) and find a solution: 
\begin{eqnarray}
& \kappa = - 0.0103 + 0.000606 i \;, \nonumber\\ 
& |c_d| = 6.32  \; . & 
\end{eqnarray}
Once these parameters, $|c_d|$ and $\kappa$, are determined, we can describe 
all the fermion mass matrices as a functions of $\sigma$ from the mass matrix 
formulas of Eqs.~(\ref{massmatrix}), (\ref{M10}) and (\ref{M126}). 
Thus in the minimal SO(10) model we have almost unambiguous Dirac neutrino 
mass matrix $m_D$ and, therefore, we can obtain the informations on $M_D$ 
from the neutrino experiments via 
$M_{\nu}=(M_D^{-1} m_D)^{T} \mu_s (M_D^{-1}m_D)$ as in Eq.~(\ref{6}). 

Now we proceed to the numerical calculation of $M_D$ from 
the well-confirmed neutrino oscillation data. The MNS mixing matrix $U$ 
in the standard parametrization is 
\be
U=
\left(
\begin{array}{ccc}
c_{13}c_{12} & c_{13}s_{12}e^{i\varphi_2} & s_{13}e^{i(\varphi_1-\delta)} \\
(-c_{23}s_{12}-s_{23}c_{12}s_{13} e^{i\delta})e^{-i\varphi_2}
&c_{23}c_{12}-s_{23}s_{12}s_{13} e^{i\delta} 
&s_{23}c_{13} e^{i(\varphi_1-\varphi_2)} \\
(s_{23}s_{12}-c_{23}c_{12}s_{13} e^{i\delta})e^{-i\varphi_1}
 & (-s_{23}c_{12}-c_{23}s_{12}s_{13} e^{i\delta})
e^{-i(\varphi_1-\varphi_2)} 
& c_{23}c_{13} \\
\end{array}
\right)\;,
\label{MNS}
\ee
where $s_{ij}:= \sin \theta_{ij}$, $c_{ij}:=\cos \theta_{ij}$ 
and $\delta$, $\varphi_1$, $\varphi_2$ are the Dirac phase and 
the Majorana phases \cite{Majorana-phases}, respectively. 
Recent KamLAND data tells us that 
\footnote{Our convention is $\Delta m_{ij}^2 = m_i^2 -m_j^2$.}
\bea
\Delta m^2_{\oplus} &=& 
\Delta m^2_{32} 
\ =\ 2.1 \times 10^{-3}\;\; {\mathrm{eV}}^2\;,
\nonumber\\
\sin^2 \theta_{\oplus} &=& 0.5\;,
\nonumber\\
\Delta m^2_{\odot} &=& 
\left| \Delta m^2_{21} \right| 
\ =\ 8.3 \times 10^{-5}\;\; {\mathrm{eV}}^2\;,
\nonumber\\
\sin^2 \theta_{\odot} &=& 0.28\;,
\nonumber\\
|U_{e3}|^2 &<& 0.061\;. \label{eqs030401}
\eea
For simplicity we take $U_{e3}=0$. Note that we can take both signs 
of $\Delta m^2_{21}$, $\Delta m^2_{21} > 0$ or $\Delta m^2_{21} < 0$.  
The former is called normal hierarchy, the latter is called inverted 
hierarchy. Here we adopt the former case, and take the lightest neutrino 
mass eigenvalue as $m_\ell = 10^{-3} \;{\rm [eV]}$. 
Then the mass eigenvalues are written as 
\bea
m_1 &=& m_\ell\;,
\nonumber\\
m_2 &=& \sqrt{m_\ell^2 + \Delta m^2_{\oplus}}\;,
\nonumber\\
m_3 &=& \sqrt{m_\ell^2 + \Delta m^2_{\oplus} + \Delta m^2_{\odot}}\;.
\eea
For the light Dirac neutrino mass matrix $m_D$, we input the SO(10) 
predicted one as was done in the previous section. 
However, unlike the case of minimal SO(10) GUT model, we can not fix $\sigma$.
(the only unknown parameter in the minimal SO(10) model before 
fitting with neutrino oscillation data \cite{Matsuda:2000zp}).
So we can obtain the heavy Dirac neutrino mass matrix $M_D$ as 
a function of $\mu_s$ and the three undetermined parameters, 
$\sigma$, two Majorana phases $\varphi_1$ and $\varphi_2$ 
in the MNS mixng matrix for fixed $U_{e3} = 0$. 
We note that the Dirac phase has little effect on our calculations 
if $U_{e3}$ has non-zero tiny values. 

\begin{figure}[p]
\begin{center}
\includegraphics[width=.8\textwidth]{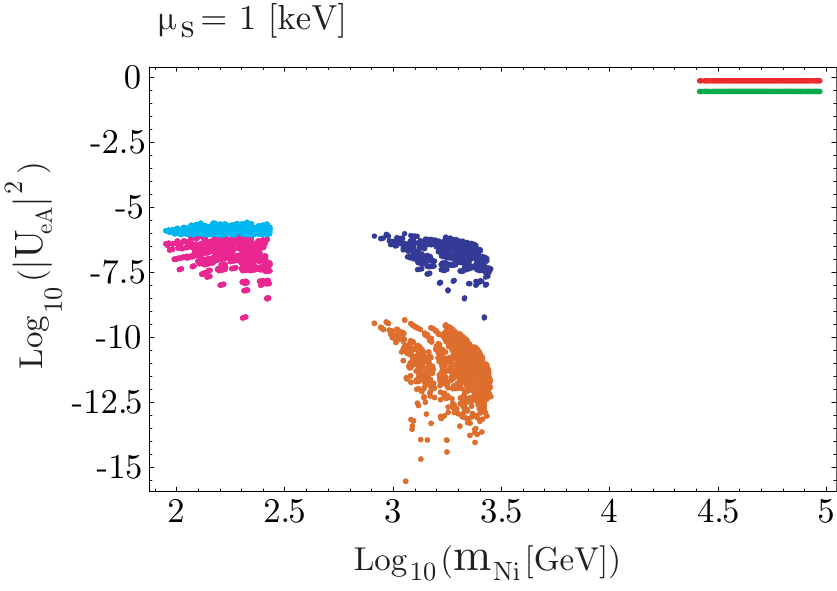}
\end{center}
\caption{Constraint on the heavy sterile mixing angles in cases
of $\mu_s = 1$ [keV] with varied $\varphi_1$, $\varphi_2$ 
and $\sigma$ from $0$ to $2 \pi$.}
\label{Fig1}
\end{figure}

\begin{figure}[p]
\begin{center}
\includegraphics[width=.8\textwidth]{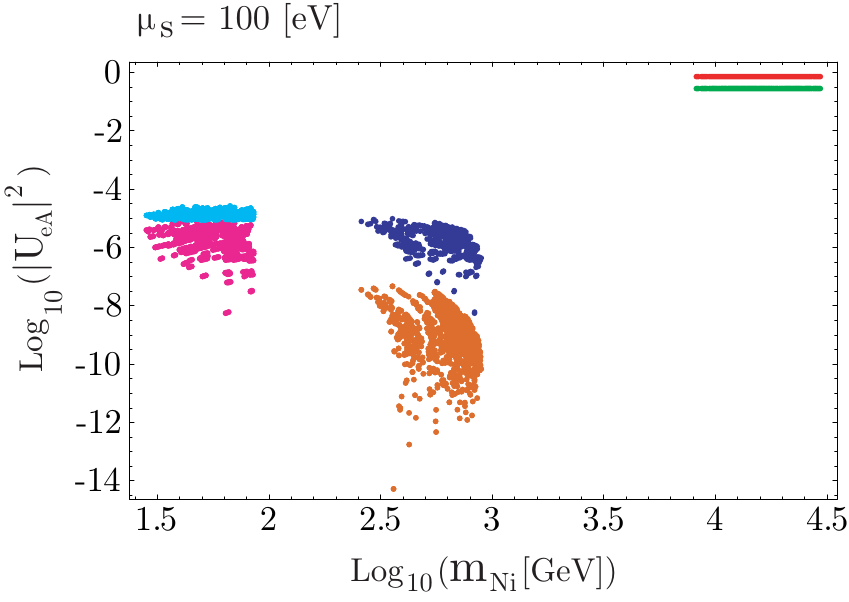}
\end{center}
\caption{Constraint on the heavy sterile mixing angles in cases
of $\mu_s = 100$ [eV] with varied $\varphi_1$, $\varphi_2$ 
and $\sigma$ from $0$ to $2 \pi$.}
\label{Fig2}
\end{figure}

\begin{figure}[p]
\begin{center}
\includegraphics[width=.8\textwidth]{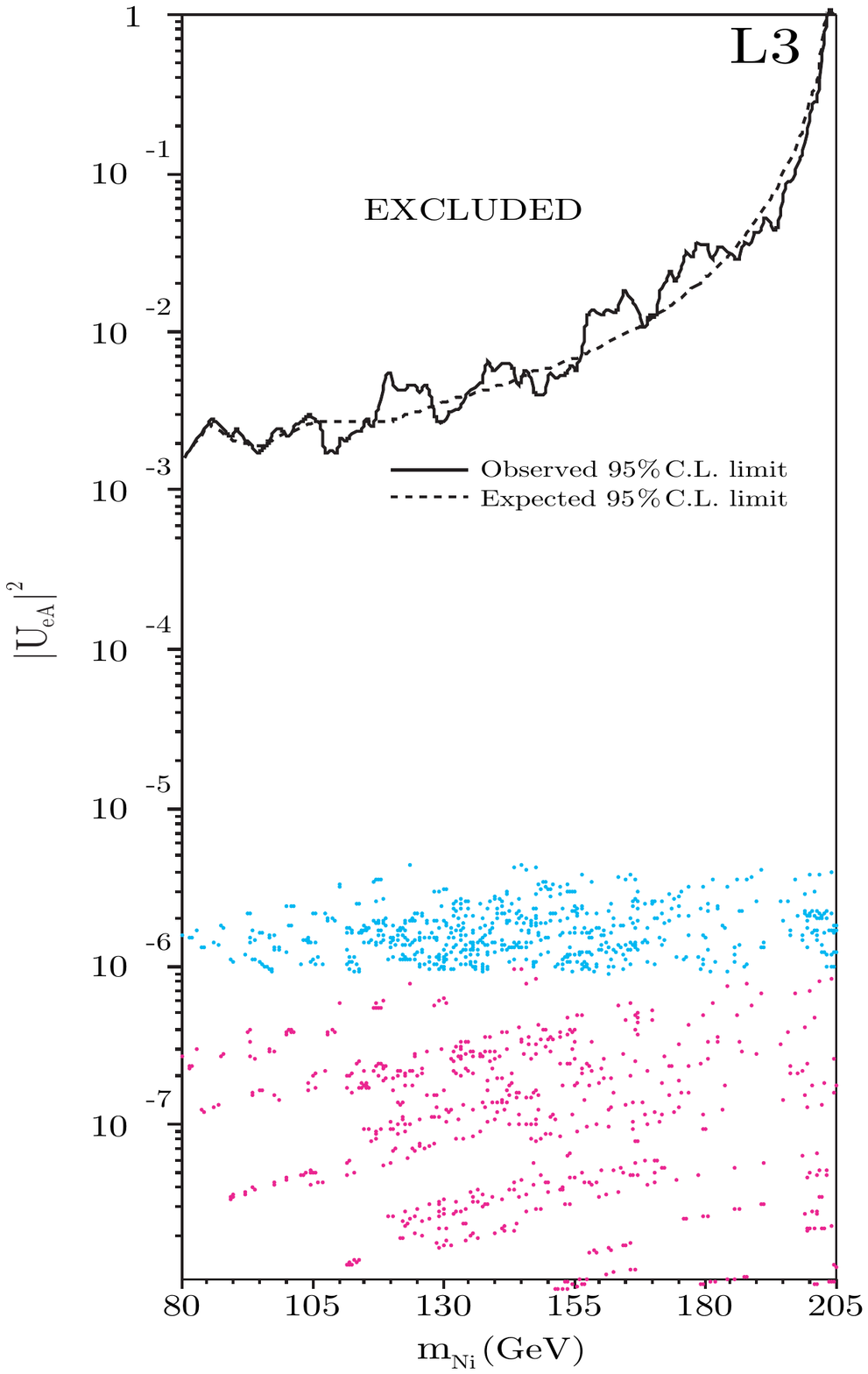}
\end{center}
\caption{The same figure as Fig.~\ref{Fig1} but
is overwritten in the LEP experimental bound.}
\label{Fig3}
\end{figure}

\begin{figure}[p]
\begin{center}
\includegraphics[width=.8\textwidth]{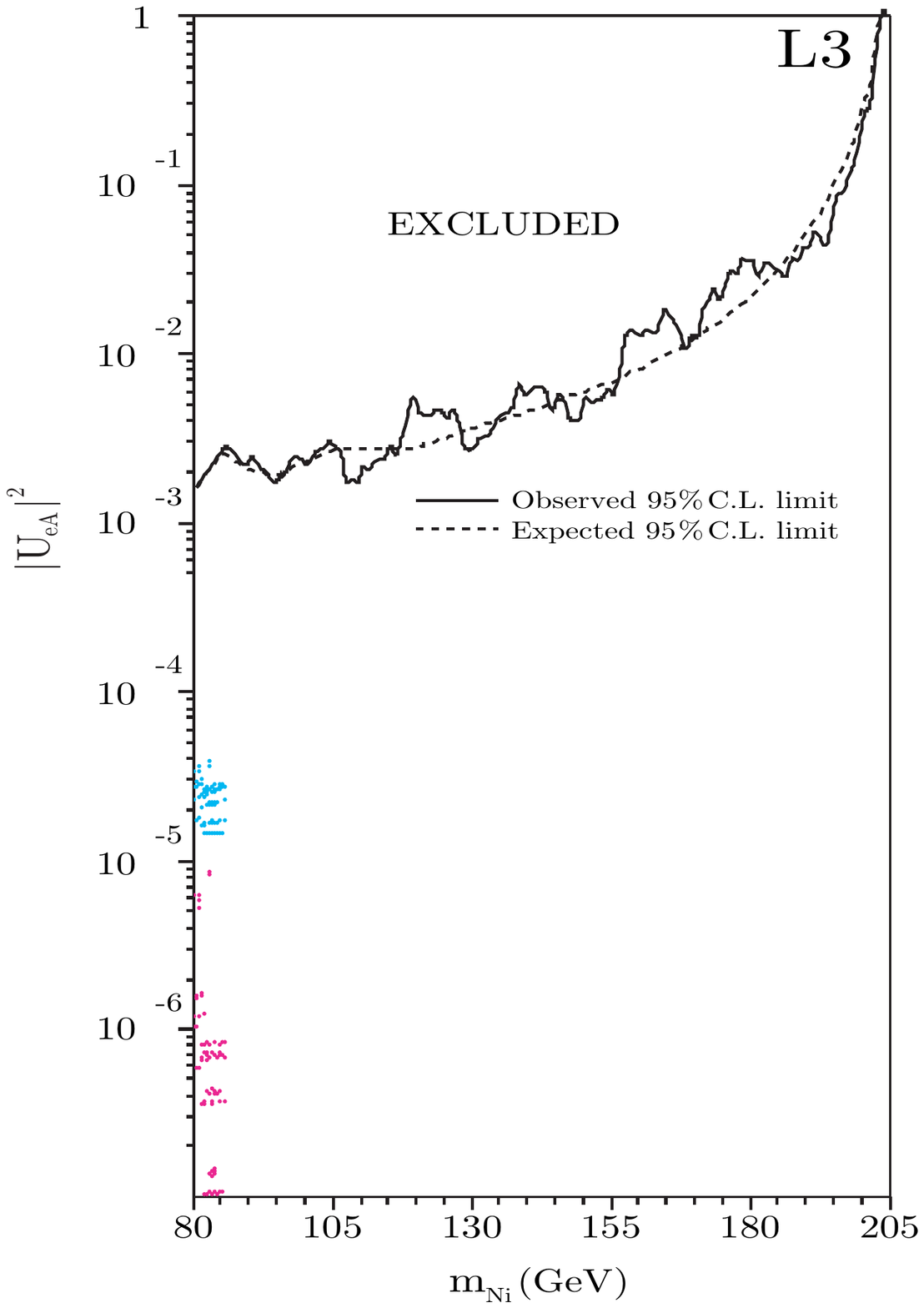}
\end{center}
\caption{The same figure as Fig.~\ref{Fig2} but
is overwritten in the LEP experimental bound.}
\label{Fig4}
\end{figure}

Then, we get a prediction 
on the mass spectra and the active to sterile neutrino mixing angles 
for $\mu_s = 1$ [keV] in Fig.~\ref{Fig1} and \ref{Fig2}.
In these Figures we varied the parameters $\varphi_1$, $\varphi_2$ 
and $\sigma$ from $0$ to $2\pi$. The same results for the case of 
$\mu_s = 100$ [eV] are shown in Fig.~\ref{Fig3} and \ref{Fig4}.
This shows that if the parameter $\mu_s$ varies from 1 [keV] to 100 [eV], 
then we obtain the result which shows one order of magnitude larger 
mixing angles and one half order of magnitude smaller mass eigenvalues. 
That is similar for the case of larger value of the parameter $\mu_s$.
These results of Fig.~\ref{Fig3} and \ref{Fig4} show that there exist
a parameter space, which is allowed by the LEP experimental bound
\cite{Achard:2001qv}.
The allowed ranges for each mass eigenvalues and the mixing angles
are listed in Table \ref{table1} and \ref{table2}.

Also it may be worthwhile noticing that such keV scale lepton number 
violation may lead to an interesting signature in the neutrinoless double 
beta decay \cite{Berezhiani:1992cd} or becomes a possible candidate for 
the cold dark matter \cite{Berezinsky:1993fm}.
These subjects are the topics for the future study.

Finally, it is remarkable to say that the see-saw mechanism itself 
(or the types of it) can never been proofed and all the models should 
take care of all the types of the see-saw mechanism including 
the alternatives to it \cite{Murayama:2004me, Smirnov:2004hs}. 
The test of all these models is due to the applications to the other 
phenomelogical consequences, for example, the lepton flavour violating 
processes and so on \cite{Deppisch:2004fa, Ilakovac:1994kj}. 

\section{Summary}
In this paper, we have constructed an SO(10) model in which the smallness 
of the neutrino masses are explained in terms of the double see-saw 
mechanism. To evaluate the parameters related to the singlet neutrinos, 
we have used the minimal SUSY SO(10) model. This model can simultaneously 
accommodate all the observed quark-lepton mass matrix data with appropriately 
fixed free parameters. Especially, the neutrino-Dirac-Yukawa coupling 
matrix are completely determined. Using this Yukawa coupling matrix, 
we have calculated the masses and mixings for the not-so-heavy singlet 
neutrinos. The obtained ranges of the mass of $M_D$ is interesting since 
they are potentially testable by a forthcoming LHC experiment. 

\section*{Acknowledgments}
The authors thank the Yukawa Institute for Theoretical Physics at
Kyoto University. Discussions during the YITP workshop
YITP-W-05-02 on "Progress in Particle Physics 2005" were useful to
complete this work.
The work of T.F. was supported in part by the Grant-in-Aid for 
Scientific Research from the Ministry of Education, Science and Culture 
of Japan (\#16540269).
T.K. would like to thank K.S. Babu
 for his hospitality at Oklahoma State University.
The work of T.K. is supported by the Research
Fellowship of the Japan Society for the Promotion of Science (\#1911329).

\pagestyle{empty}

\begin{table}[p]
\begin{tabular}{|c|c|}
\hline \hline
The allowed ranges for mass eigenvalues 
& The allowed ranges for mixing angles \\
\hline
$89.332~{\rm [GeV]} < m_{N_1} < 270.13~{\rm [GeV]}$ & 
$ -6.0632 < \log_{10} \left(|U_{e4}|^2 \right) < -5.5840 $ \\
$89.332~{\rm [GeV]} < m_{N_2} < 270.13~{\rm [GeV]}$ & 
$ -9.2564 < \log_{10} \left(|U_{e5}|^2 \right) < -6.0151 $ \\
$819.72~{\rm [GeV]} < m_{N_3} < 2.8259~{\rm [TeV]}$ & 
$ -9.2564 < \log_{10} \left(|U_{e6}|^2 \right) < -6.0151 $ \\
$819.72~{\rm [GeV]} < m_{N_4} < 2.8259~{\rm [TeV]}$ & 
$ -15.530 < \log_{10} \left(|U_{e7}|^2 \right) < -9.3336 $ \\
$25.988~{\rm [TeV]} < m_{N_5} < 93.410~{\rm [TeV]}$ & 
$ -0.55515 < \log_{10} \left(|U_{e8}|^2 \right) < -0.55064 $ \\
$25.988~{\rm [TeV]} < m_{N_6} < 93.410~{\rm [TeV]}$ & 
$ -0.14498 < \log_{10} \left(|U_{e9}|^2 \right) < -0.14047 $ \\
\hline \hline
\end{tabular}
\caption{The allowed ranges for mass eigenvalues and mixing angles
in case of $\mu_s = 1$ [keV].}
\label{table1}
\end{table}
\begin{table}[p]
\begin{tabular}{|c|c|}
\hline \hline
The allowed ranges for mass eigenvalues 
& The allowed ranges for mixing angles \\
\hline
$28.254~{\rm [GeV]} < m_{N_1} < 85.443~{\rm [GeV]}$ & 
$ -5.0632 < \log_{10} \left(|U_{e4}|^2 \right) < -4.5841 $ \\
$28.254~{\rm [GeV]} < m_{N_2} < 85.443~{\rm [GeV]}$ & 
$ -8.2575 < \log_{10} \left(|U_{e5}|^2 \right) < -5.0161 $ \\
$259.26~{\rm [GeV]} < m_{N_3} < 893.68~{\rm [GeV]}$ & 
$ -8.2575 < \log_{10} \left(|U_{e6}|^2 \right) < -5.0161 $ \\
$259.26~{\rm [GeV]} < m_{N_4} < 893.68~{\rm [GeV]}$ & 
$ -14.284 < \log_{10} \left(|U_{e7}|^2 \right) < -7.3339 $ \\
$8.2182~{\rm [TeV]} < m_{N_5} < 29.540~{\rm [TeV]}$ & 
$ -0.55294 < \log_{10} \left(|U_{e8}|^2 \right) < -0.55285 $ \\
$8.2182~{\rm [TeV]} < m_{N_6} < 29.540~{\rm [TeV]}$ & 
$ -0.14268 < \log_{10} \left(|U_{e9}|^2 \right) < -0.14267 $ \\
\hline \hline
\end{tabular}
\caption{The same table as Table.~\ref{table1} 
but in case of $\mu_s = 100$ [eV].}
\label{table2}
\end{table}

\newpage

\end{document}